# A Speed Limit for Evolution: Postscript

Robert Worden

Wellcome Centre for Human Neuroimaging, Institute of Neurology, University College London, London, United Kingdom

rpworden@me.com

September 2022

Abstract:

In 1995 I wrote a paper 'A Speed Limit for Evolution' – whose main result was that evolution must proceed rather slowly, in accordance with the earlier views and intuitions of many authors. The abstract of the paper said: '*An upper bound on the speed of evolution is derived. The bound concerns the amount of genetic information which is expressed in observable ways in various aspects of the phenotype. The genetic information expressed in some part of the phenotype of a species cannot increase faster than a given rate, determined by the selection pressure on that part. This rate is typically a small fraction of a bit per generation. Total expressed genetic information cannot increase faster than a species-specific rate—typically a few bits per generation.*' This result was derived in the presence of sexual reproduction and other effects such as temporarily isolated sub-populations.

In 1999 David Mackay published a paper which apparently contradicted this result. In the abstract, he wrote '*We study a simple model of a reproducing population of N individuals with a genome of size G bits. …. We find striking differences between populations that have recombination and populations that do not. If variation is produced by mutation alone, then the entire population gains up to roughly 1 bit per generation. If variation is created by recombination, the population can gain of the order of $\sqrt{G}$ bits per generation.*' So Mackay proposed that there were outstanding evolutionary benefits to sexual reproduction, and that my previous result was too low by a very large factor, up to about $10^4$ for *homo sapiens*. He later repeated this result in a textbook he wrote in 2003.

The purpose of this note is to show that the key assumption of Mackay's model, that '*fitness is a strictly additive trait*' is so unrealistic as to render his results irrelevant to any actual life form. In consequence, the speed limit I derived is still valid, and has important consequences for human cognitive evolution. In the past 2 million years, while the human brain has greatly expanded and our intelligence has grown, the amount of new design information in the human brain must be very small compared to the design information already in the mammalian brain.

# 1. Introduction

Darwinian evolution is one of the most successful scientific theories, but its successes lie largely in retrospective accounts of how species have evolved, rather than in contemporary measurements of the process of evolution. While there are some data about evolution in action at the present day, they are scarce - mainly because it takes a long time for many evolutionary changes to occur.

While many authors believe, from evidence and from a common-sense model of how evolution happens, that evolution is a slow process, it is another matter to show by mathematical analysis that evolution **must** be a slow process. It is difficult because of the large (and ill understood) complexity of the mapping between genotype and phenotype, because of the complexities of sexual reproduction, finite and inhomogeneous populations, and other real-world complications.

In [Worden 1995] I derived an upper bound on the speed of evolution, in the presence of some of these complexities – notably sexual reproduction. I published the result in the Journal of Theoretical Biology. The paper was refereed by John Maynard Smith.

In many ways the result was unremarkable, and simply confirmed what many people had believed all along – that the total amount of useful design information, expressed in the genotype of any species, can only increase by of the order of one bit per generation. Intuitively, differential survival only feeds information into the genotype of species at approximately this rate.

In [Mackay 1999, 2003] David Mackay published a result which apparently contradicted this result. In the abstract of [Mackay 1999], he wrote '*We study a simple model of a reproducing population of N individuals with a genome of size G bits: fitness is a strictly additive trait subjected to directional selection; variation is produced by mutation or by recombination and truncation selection selects the N fittest children at each generation to be the parents of the next. We find striking differences between populations that have recombination and populations that do not. If variation is produced by mutation alone, then the entire population gains up to roughly 1 bit per generation. If variation is created by recombination, the population can gain of the order of √G bits per generation.*' He later repeated the result in a textbook he wrote [Mackay 2003].

This result contradicted my previous result, and could contradict it by a large amount. In the case of the human genome (of the order of $10^8$ bits of information), instead of crawling along at about one bit per generation (as I and others had supposed), evolution could zip along, at up to $10^4$ bits of useful new design information per generation. At this speed, mankind could have truly become 'what a piece of work' – noble in reason, infinite in faculty, and all the rest.

Mackay's result, if applicable to actual biological species, would have enormous repercussions – in making sexual reproduction massively favoured over asexual reproduction, and in enabling any species to adapt very rapidly to changes in its habitat. If Mackay's limit were true, evolutionary arms races could take place at warp speed.

However, Mackay's result is not relevant to any biological system, for two important reasons.

1. He does not define any concept of genetic information expressed in the phenotype - which is what matters in evolution, and is what biologists study. He relies instead on some poorly-defined concept of 'fit' information in the genome; a concept which is so imprecise as to be meaningless.
2. He relies on a central assumption – that fitness is an additive function of fitness contributions from different genes – which is so unrealistic that his model is not relevant to any existing species.

These points are elaborated in this paper. The result is that Mackay's results are of no biological interest, and do not contradict my earlier speed limit.

In the last section of this paper, I revisit some results of my earlier paper [Worden 1995] for human cognitive evolution – which are still important for ongoing discussions of the evolution of human intelligence and language.

Researchers and the public are rightly concerned that science sometimes (or often!) goes wrong; so that when it does go wrong, it is important to correct the errors. Science is nothing if it is not a self-correcting system. David Mackay was a highly respected scientist, and made important contributions in several fields. So his writings are influential; and when, as in this case, he made an uncharacteristic error, it is important to correct it. When David Mackay wrote his paper in 1999, I did not have the time to reply properly to it. He died in 2016, as is so no longer able to defend his results. If anyone else would like to do so, I would be happy to debate it with them.

# 2. The Speed Limit for Evolution

This section briefly recapitulates the key points of my 1995 speed limit paper.

In order to derive any speed limit for evolution, we need some measure of the amount of useful design information expressed in the phenotype of a species. I defined a measure which I called **Genetic Information in the Phenotype** (GIP).

GIP is a relative measure of information (not an absolute one[1]) defined not for any individual in the species, but for a species as a whole. The simplest way to understand GIP is to approach it through examples:

1. If the individuals in species A have a random mixture of two eye colours – 50% blue and 50% brown – while the individuals of species B all have blue eyes, then in that respect, the GIP of species B is one bit larger than the GIP of species A. Species B has made one binary choice.
2. If the individuals in species A have a Gaussian distribution of weights 50Kg ±10Kg – while the individuals of species B have a distribution 50Kg ±5Kg, then in that respect, the GIP of species B is one bit larger than the GIP of species A

GIP is a Shannon-like sum of (p ln(p)) across all measurable phenotypic traits. GIP depends on probability distributions across the population of a species. It cannot be measured for any one individual. You can add in GIP contributions for any phenotypic attribute you are able to measure, organ by organ, or behaviour trait by trait. For a complex organ such as the brain, there are many parameters to vary, and so there is a large GIP. We cannot measure GIP absolutely, as we may keep finding new traits; but given any definition of GIP, we can measure the change in GIP between two generations of a species, or between two species (such as *homo* and chimpanzees).

There is no simple relationship between GIP and the information content of the genotype, which is roughly 2 bits for every base pair – and can be very large. In this sense, the genetic encoding of form and behaviour is a very inefficient encoding. Changes in raw genetic information cannot be equated with changes in GIP.

GIP can change over the generations, by processes of differential survival, in the absence of any mutations or recombination. For instance in the example above, if blue eyes survive 20% better than brown eyes, then over many generations an A-like population, with a 50:50 mix of two eye colours, will change to a B-like population of pure blue eyes – gaining one bit of GIP in the process.

The speed limit I derived in 1995 is then of the following form: if a species has an average of N offspring per mating pair, then the maximum long-term rate of increase of GIP is less than approximately $\log_2(N/2)$ bits per generation. This total increase in GIP must be partitioned into positive increases of GIP for different attributes of the phenotype (intelligence, immune system, physical attributes, etc.). So, for instance, we expect the maximum increase in GIP in the design of the brain to be less than 1 bit per generation.

If there is only asexual reproduction, with no mutations, then from that process, it is easy to show that the rate of change of GIP is subject to the speed limit. Harmful mutations will make the rate of increase in GIP slower than this, and the need to find beneficial mutations will require more time, making the process yet slower – and so, still subject to the same upper limit.

The bulk of my 1995 paper was devoted to showing that the same speed limit still holds in the presence of some real-world complications – notably, sexual reproduction with recombination, and inhomogeneous and finite populations. The mathematics were not rigorous - leaving open the possibility that some complexity that I had overlooked might upset the result – but, I expected, by small amounts which would be overwhelmed by the many other slowing-down factors, notably the time needed to discover beneficial innovations.

The speed limit applies only to the long-term average rate of increase of GIP, and can be violated in the short term. An example is where some complex organ or behaviour pattern is lost in evolution, and then rediscovered. Consider moles, which do not need eyes. If the embryonic development of mole eyes had been completely switched off by some gene – so that eyes never appear in the mole phenotype – yet the genetic specification of the eye was retained in their genes, it would be possible for a mole descendant species to emerge into the light and 'rediscover' eyes within a small number of generations – recovering all the GIP for a complex organ in a short space of time, and violating the speed limit over that period. The longer-term average, however, is constrained by the speed limit (assuming that most evolution consists of innovation rather than re-discovery).

### 3. Mackay's Model of Evolution

Mackay says early in his paper [Mackay 1999] *'we choose a crude model because it readily yields simple but striking results'*. He then states the key assumption of his model, his assumption of additive fitness:

*'The genotype of each individual is a vector **x** of G bits, each having a good state $x_g = 1$ and a bad state $x_g = 0$. The fitness $F(\mathbf{x})$ of an individual is simply the sum of her bits, $F(\mathbf{x}) = \sum x_g$. The bits in genome can be considered to correspond either to good alleles ($x_g = 1$) and bad alleles ($x_g = 0$) or to the nucleotides of a genome. We will concentrate on the latter interpretation. The essential property of fitness that we are assuming is that it is locally a roughly linear function of the genome, that is, that there are many possible changes that one could make to the genome, each of which has*

---

[1] In just the same way as GIP, physical positions and times cannot be defined absolutely, but are only defined relative to other positions and times. Nevertheless they are very useful quantities.

*a small effect on fitness, and that these effects combine approximately linearly.'*

Given this assumption, the results of his model follow straightforwardly, from simple computer simulations or mathematical analysis.

In his model of recombination, the population doubles by reproduction, and halves by death before reproduction, in each generation. Only the fitter half of the population go on to reproduce. In the first generation, the bits are set at random, so the fitness of individuals has a mean $F = G/2$ and a standard deviation of the order of $\sqrt{G}$. In the next generation, each child receives roughly half its genotype from one parent, half from the other parent (the proportions vary, with a variable crossing point). The standard deviation of each half of the childrens' genotype is still of the order $\sqrt{G}$, so the standard deviation of the whole next-generation genotype is also of order $\sqrt{G}$. This leads to a Gaussian distribution of children's fitness, and the bottom half of the Gaussian is removed by early deaths of the less fit half of the population – leading to a half-Gaussian distribution, with zero proabability density below the previous mean. So the mean fitness increases by an amount of the same order as the standard deviation, of the order $\sqrt{G}$. This carries on over many generations, increasing the average fitness by an amount of order $\sqrt{G}$ each generation. It is that simple.

Mackay then simulates his model for finite populations, and for non-uniform probabilities of crossovers, and retreating slightly from the additive model of fitness – to a nearly additive model, where the fitness depends on sums of pairwise products for parts of the genome- which is still an essentially local additive model. In all cases, his fast increase result, that fitness increases by the order of $\sqrt{G}$ per generation, is unchanged.

Discussing his results, Mackay says first:

"*The rate at which information can be acquired by natural selection was addressed by Kimura (1961). Using two different arguments, he asserts that the maximum rate of information acquisition is 1 bit per generation (assuming that a generation is the time taken for the population to double).*"

He then says:

"*That the 'speed limit for evolution' is about one bit per generation is also 'proved' by Worden (1995), who goes beyond Kimura by showing that this result holds whether or not the species has sex. This contrasting conclusion is reached because of Worden's choice of model. Instead of defining an individual's fitness as a relative quantity, involving competition with other individuals, his model assumes that one's genotype determines one's probability of having children absolutely.*"

I agree with Mackay on two points: first, that my 'proof' may well deserve his scare quotes, as not being a rigorous mathematical proof; and second, that both of our results depend on our choice of model. However, I assert that the issue of competitive versus absolute reproductive fitness is a red herring, having little effect on the results of my model; and that my assumptions are designed to be biologically realistic as far as possible; whereas Mackay's 'additive fitness' assumption is absurd – so unrealistic, that his results are not relevant to any living species.

## 4. Mackay's Definition of Information

In order to discuss the rate of increase of information through evolution, you first need to define the information. This may seem a simple matter, but it needs some consideration. My 1995 paper hinged on the idea of 'Generic Information in the Phenotype', or GIP. The definition of GIP depends on a species population, and corresponds intuitively to the total information content expressed in an individual's cells, body plan and behaviour. It relates to biologically measurable properties, which drive evolutionary changes.

Mackay does not define or use any concept of phenotypic information – but uses a concept of information in the genotype, which is never defined, and which on close inspection, turns out to be undefinable.

For instance, in one model he considers the genotype to be a sequence of G locations, each of location having two possible bases. On the face of it, then, you might expect its information content to be exactly G bits, which does not change as the bases change – leading to a zero rate of change. He avoids this zero by considering each of the G locations to have either a 'fit' value of an 'unfit' value – and taking the information content to be the sum of the fit values. This measure does not have a useful relation to either the genotype or the phenotype. It does not have any relation to the genotype, except to a privileged observer who can distinguish the 'fit' bits from the 'unfit' bits – which no observer can do. In any case, the amount of raw information in the genotype can be many billions of bits, and bears no useful relation to the amount of design information in the phenotype.

So not having defined any useful measure of genetic information, it is already evident that Mackay's results cannot contradict my previous results, as he implies they do. There is a second reason why they do not – because of his assumed additive model of fitness.

## 5. The Assumption of Additive Fitness

Mackay presents his assumption of additive fitness in two different versions. I first address Mackay's second interpretation of his assumption that fitness is an additive function of the choices of nucleotides in an animal's DNA. His 'Version 2' assumption is that fitness depends additively on the nucleotides in a genome.

For any species, fitness depends on having working proteins, which selectively catalyse and control many chemical reactions in the cell. Nobody would assert that the folding of a protein – which determines its enzymatic action – is a linear function of its sequence of amino acids; that is, a linear function of a sequence of DNA bases. The folding of a protein is a highly complex consequence of its amino acid sequence – a problem which has only recently been solved by the most advanced AI techniques.

To assert that a linear function will solve this problem is a massive over-simplification, which contradicts a huge scientific consensus; yet this is what Mackay's 'version 2' assumption implies. It implies that the fitness of a protein – its ability to selectively catalyse a particular chemical reaction - can be progressively tweaked up from 0 to 1, by independently switching base pairs, because its effectiveness only depends on a sum of 0s and 1s for each DNA base pair of its gene. So, for instance, you could swap the position of any 1 with the position of any 0, with no effect on the catalytic activity of the protein. Such an assumption would make the evolutionary search for working proteins into a trivial matter, and is a travesty of modern chemistry and biology.

I next turn to version 1 of Mackay's additive assumption – that fitness is just a sum of 0s and 1s for 'good' and 'bad' alleles. Even in the simplest cases, this is unrealistic. Consider any simple chain or cycle of chemical reactions in a cell, such as the Krebs cycle. The effectiveness of the chain depends on all the different reactions working – each one being catalysed and controlled by its own enzymes. Fitness is at the very least a multiplicative function of the genes coding for the different enzymes. Additivity of enzyme fitness is the most unrealistic possible model.

The problem is yet more acute for the design of any complex organ such as the eye or the brain. For any complex design to work, many aspects of the design must fit together in a coordinated way. Nerve cells of type A must connect to cells of type B, but not to type C or D; the connections must have certain properties in order to make the right computation, and so on. In other words, any complex organ, or even a part of it, is a high-wire act, where many separate pieces of GIP must fit together correctly to work. If one detail is wrong, the organ does not work. Again, fitness is like a product of many different terms; if one term is zero, the whole product is zero. This is the opposite of Mackay's additive model of 'throw in enough good ingredients to get a good result'.

Furthermore, in the 'good allele/bad allele' version of Mackay's additive assumption, we would expect the bad alleles to vastly outnumber the good alleles. It is not just a matter of 'turning on' a good allele; many generations are required to discover a good allele in the first place, further reducing the average speed of evolution.

Finally, Mackay's additive assumption (in either form) implies either a massive reduction in the effective information content of the genome – from G bits down to $\ln_2(G)$ bits – or some amazing set of coincidences, in that many distinct phenotypes all have identical fitness. This is because in his model, there are only G different levels of fitness, integers from F = 0 to F = G. If each level corresponds to one phenotype, the genotype is only encoding $\ln_2(G)$ bits of information about the phenotype. If each level corresponds to many different phenotypes, allowing up to $2^G$ phenotypes in total, this requires some amazing (and vanishingly unlikely) coincidences, that of the order of $(2^G/G)$ distinct phenotypes all have the same fitness, for each value of F from 0 up to G. Either option, or anything in between them, fails to be biologically realistic.

For these reasons, all possible versions of Mackay's additive fitness assumption are obviously unrealistic, and cannot be relevant to any living species. His 'fast evolution' result, which contradicts my previous speed limit, does not do so in any biologically relevant cases. The previous speed limit stands.

## 6. Counter-Examples to Mackay's Model

I have tried to find illustrations of what Mackay's speed limit would imply, to illustrate its absurdity. This effort has been only partly successful, for reasons which may be seen below. Nevertheless, it does illustrate the kinds of intellectual contortions that would be needed to believe Mackay's result.

As an example, I consider some fairly complex evolutionary innovation, which requires, say, 40 bits of new GIP in order to deliver a 20% increase in fitness (where fitness is defined as the expected number of offspring per individual). The 40 bits of GIP might consist of a new sense organ (such as a detector of electric fields in a fish), or a new behaviour pattern to evade predators.

We consider any evolutionary innovation to emerge in two phases:

1. The emergence of the new phenotype in some small proportion of the population – as small as one individual.
2. The proportion of the population with the new phenotype grows to be close to 1.0, because of its greater fitness.

Because the new phenotype contains 40 bits of design information, this implies that the proportion of the population which have it by chance is of the order of $2^{-40}$, or about $10^{-12}$. This means that for realistic populations over many generations, there is almost no chance of any individual ever having the right set of traits, so the process does not get started. So the set of traits must evolve incrementally – for instance in 10 stages of 4 bits each, each stage giving a fitness increase of 2%.

We can compute the evolutionary time required for the first of the 10 stages – in particular, for step (2) of that stage.

Having 4 bits of GIP correctly set at random implies that about one in $2^4 = 16$ individuals have the required phenotype at random. The proportion in the population will increase from 1 in 16, by a factor of 1.02 per generation (2% extra fitness), so it will need approximately 130 generations to grow to half the population having the first-stage change – at which time the second of the ten stages can effectively begin. So the whole evolution of 10 stages requires of the order of 1300 generations. The rate of increase of GIP is $40/1300 = 0.03$ bits per generation. This is much less than the limit of 1 bit per generation, partly because the selection pressure of 20% is less than the maximum possible pressure.

Note that this result only depends on adding the time required for stage (2) at each incremental step toward the final state. No time has been added for stage (1) – and if it were added, it would make the rate of evolutionary change yet slower.

How, then, could sexual reproduction and recombination bring about the great acceleration of the rate of evolution, as required by Mackay's much higher speed limit? Although others may be able to, I can think of no mechanism to do it. The main reason is that while sexual reproduction might somehow accelerate stage (1) – the emergence of novelty in a small part of the population - there seems to be no way for it to accelerate stage (2) – the growth of a new phenotype to dominate a population. Stage (2) on its own leads to the low rate of evolution; and the rate of stage (2) depends only on simple arithmetic, not on any subtlety of sexual/asexual reproduction.

This is not a rigorous argument, and others might find holes in it. But it does illustrate that even if evolution consists only of fitter variants displacing less fit variants, it takes a long time – to which must be added the time in which the fitter variants are created.

For typical genotype sizes of $G = 10^6$ or more, the Mackay speed limit implies an acceleration of the order of 1000 – to 1000 bits per generation. In his model, the evolution of a new 40-bit trait, as in the example above, could have easily happened within one generation. How? I can find no plausible model of evolution in which this could happen. This further illustrates that Mackay's limit is not relevant to any actual species.

## 7. Consequences for Cognitive Evolution

Animal brains have evolved over a period of more than 500 million years. We know this because 540 million years ago, trilobites had complex eyes, with many thousands of receptors; there would have been no point in gathering all this information without a well evolved brain to process it. Trilobites were soon followed by the Cambrian explosion – a huge diversification of body plans and behaviour, which must have required changes in brain design.

In this time of half a billion years, animals have gone through an estimated 100 million generations, undergoing selection pressure from many different causes such as starvation, predation, and disease. At each generation the total increase in GIP, in all aspects of body design and behaviour, was at most of the order of 1 bit. The GIP increase from a more evolved CNS could only be a part of the total increase in GIP – say for illustration, 1/10 of a bit per generation, or 10 million bits. This is an approximate upper limit on the design complexity of any animal brain.

How much extra design information might there be in the human brain, compared to other animal brains? The period of rapid enlargement of the human brain has only lasted approximately 2 million years [DeSilva et al 2021], which is less than 1% of the total time (in years, or in generations) over which animal brains have evolved. If we assume, consistent with the speed limit, that the total increase in human GIP over this period has been limited to a few bits per generation, and that only a portion of this GIP increase has been in the brain, then it follows that the 'human brain extra' – the design complexity of human brains, compared to other primate brains – is at most only a few percent. We share about 99% of our brain design with other primates. The increase in human brain size is mainly an increase in size and computational power, with very little new brain design – reflecting more neural circuits, rather than different neural circuits.

This result has consequences for the recent evolution of human intelligence and language. It implies, in brief, that language cannot have required large amounts of new brain design, but must have been built mainly on pre-existing cognitive facilities.

One account of the evolution of language can therefore be dismissed – the account advanced by [Bolhuis, Chomsky et al. 2014]. In this account, the key innovation required for productive human language was the 'merge' operation, which enables recursive grammar, and is required to express a 'discrete infinity' of possible sentence meanings. On the account of Bolhuis et al., the evolution of the merge operation occurred within a very short timescale: "*the language faculty is an extremely recent acquisition in our lineage, and it was acquired not in the context of slow, gradual modification of preexisting systems under natural selection but in a single, rapid, emergent event that built upon those prior systems but was not predicted by them.*".

While Bolhuis et al. say that the change required to introduce the merge operation was '*relatively minor*', it was nevertheless a '*single, rapid, emergent event*' occurring less than 200,000 years ago. Since the breeding population of *homo sapiens* at this time was of order 10,000 [Harpending et al 1998], or $2^{13}$ people, the chances of a single evolutionary event in a single individual producing as much as, say, 24 bits of appropriate new design information in the brain are of the order of $2^{13} * 2^{-24} = 2^{-11}$. If an evolutionary account requires any event with probability less than one in a billion, it can safely be dismissed. On the other hand, if the innovation for the merge operation required less than 24 bits of GIP, that would be remarkably little new brain design for the key operation of merge, which has been so important for the human species. So all versions of this account can be rejected.

Ironically, the 'super-fast evolution' proposed by [Mackay 1999] would be consistent with the Bolhuis et al. account, of large amounts of useful new design information arriving in a single evolutionary event. However, as described in this paper, Mackay's result is wrong, and is not applicable to any extant species, including mankind. So it cannot be used to rescue the Bolhuis et al. account of the evolution of language.

An account of the evolution of language which is consistent with the speed limit is the account of [Worden 2022], in which language emerged over a period of up to 2 million years, through sexual and social selection, in which language was used for the display of superior intelligence. Two million years (or 10,000 generations) is sufficient time for a language faculty to evolve, consistent with the speed limit, by reuse of pre-existing primate cognitive faculties.